\documentstyle[aps,prl,twocolumn,epsfig]{revtex}

\def\up{{\uparrow}}
\def\dn{{\downarrow}}

\def\bbr{{\bf r}}
\def\bri{{{\bf r}_1}}
\def\br1{{{\bf r}_1}}
\def\br2{{{\bf r}_2}}
\def\bribr2{{\bf r}_1\!-\!{\bf r}_2}
\def\radphi{\Phi}
\def\OSFzero{\Omega^{\rm SF0}}
\def\OSF{\Omega^{\rm SF}}
\def\OOSF{\overline{\OSF}}
\def\OOSFR{\overline{\OSF_R}}
\def\OSFRzero{\OSFzero_R}

\begin{document}
\draft
\title{A method to include the spin-fluctuation
in the ab-initio electronic-structure calculation.}
\author{Takao Kotani and Hisazumi Akai}
\address{Department of Physics,
Osaka University, Toyonaka 560--0043, Japan$^1$}
\date{\today}

\maketitle
\begin{abstract}
We present a new method of ab-initio electronic-structure
calculation including the spin-fluctuation (SF) self-consistently.
We start from the Luttinger-Ward functional given
as the sum of the LDA functional plus the temperature-dependent
part of the SF energy functional.
The size of interactions used in it
are determined in a similar manner
to the self-consistent renormalization theory by Moriya and Kawabata.
Obtained paramagnetic susceptibilities on
Pd, Ni, Fe, and fcc-Co above $T_c$ show
rather good agreements with experiments.
\end{abstract}

\vspace{5mm}
\label{intro}
In magnetic metals like Fe,
LDA gives quite good description for ground state 
properties at zero-temperature $T=0$;
LDA reproduces the so-called Slater-Pauling curve
obtained by experiments very well \cite{akai1},
also does the spin waves \cite{savrasov1998}.
But neither LDA nor its crude extension to finite temperature
contains the contributions due to the spin-fluctuation (SF).
SF for magnetic materials should have strong temperature dependence,
and enforce them paramagnetic above their critical temperatures $T_{\rm c}$.
In Refs. \cite{uhl} and \cite{rosen}, methods including SF had developed
with adiabatic approximation in an ab-initio scheme;
they first calculated the adiabatic surface
of all the configuration of static SF, then calculated
the partition function assuming spin variables as classical.
It works rather well, though these approximations could be problematic
in paramagnetic phases at high $T$ where the Stonar-like excitations
and spin-wave like excitation could mix well.
In addition, the ground state, where they calculated the adiabatic surface,
can be changed at higher $T$. Especially in Pd, which is "almost-ferromagnetic",
these methods have not been applied to, presumably because
the change of ground state would cause problems.
As the opposite limit where the Heisenberg-like model works well,
we know "metallic weak-ferro-magnet", which are well described 
by the self-consistent renormalization (SCR) theory \cite{moriya} 
given by Moriya and Kawabata. They evaluate SF in a kind of 
random-phase-approximation (RPA), where we should determine the size of 
interaction in a self-consistent (SC) manner.
Here we present a new ab-initio SC method including SF,
where the size of interaction is determined 
self-consistently as in the SCR theory.
As results compared with experiments,
we show paramagnetic susceptibilities for Pd, Ni, Fe, and fcc-Co above $T_c$.
Our calculations are based on a finite-temperature version
of AkaiKKR \cite{akaikkr}, in the atomic sphere approximation (ASA).

In the treatment like LDA+$U$ and their extensions
like LDA+'dynamical mean field theory (DMFT)',
a key problem is in the determination of the effective interaction $U$.
Lichtenstein, Katsnelson, and Kotliar presented such a scheme
to describe magnetic susceptibilities of Ni and Fe at Hight $T$.
Their method can take into account higher-level of correlation effects than
our treatment, though their $U$ is given externally \cite{lichtenstein01}.
Very recently,  Biermann, Aryasetiawan, and Georges suggested
$GW$+DMFT to determine them self-consistently \cite{biermann02}.
Our method can be identified as a simplified version of such a method.



We start from the Luttinger-Ward (LW) functional $\Omega[g]$
at finite temperature. It can be written as
\begin{eqnarray}
&& \Omega[g]\!=\! {\rm Tr}\left[ \log(g) + \frac{\partial g}{\partial \tau} \right]
 + {\rm Tr}\left[ \frac{-\nabla^2}{2m} g \right]
 + E_{\rm ext}[n] \nonumber \\&&
+ E_{\rm Cou}[n]+ E_{\rm xc}[g]
\label{omegag}
\end{eqnarray} Here $g_\sigma(1,2)$ is the finite-temperature
Green function, and $1\equiv \bri\tau_1$.
$n_\sigma({\bf r})$ denotes the electron density and
is treated as the functional of $g_\sigma(1,2)$ in Eq.(\ref{omegag}).
We may identify the first term in right-had side (RHS) in Eq.(\ref{omegag})
as "the population-entropy" term
because it is reduced to be $-T$ times the entropy of the system
in the non-interacting case.
The second term is the kinetic energy term, and
$E_{\rm ext}[n]$ is the external-potential
term containing the chemical potential. Terms $E_{\rm Cou}$ and $E_{\rm xc}$
are related to the Coulomb interaction between electrons.
Sum of them are referred to as $\Phi[g]$ in literatures \cite{lw60}.
An advantage starting from the LW functional is in the conservation laws;
they apparently hold if our approximated $\Omega[g]$
keeps the corresponding symmetries; such approximations are called as
the Baym-Kadanoff approximation \cite{bktext}.
Our approximation for $E_{\rm xc}[g]$ is
\begin{equation}
E_{\rm xc}[g] \approx\ E_{\rm xc}^{\rm LDA}[n] + \overline{\Omega^{\rm SF}}[g],
\end{equation}
where the SF functional $\overline{\Omega^{\rm SF}}$ contains just
the temperature-dependent part of the SF energy written
as $\overline{\OSF} = \OSF - \OSFzero$,
where $\OSFzero$ is $\OSF$ at $T=0$ so as to remove the temperature-independent part.
We use $E_{\rm xc}^{\rm LDA}[n]$ given by \cite{vwn}.
This $\overline{\OSF}$ should have strong temperature-dependence.
The stationary-point condition
$\delta \Omega[g]/\delta g =0$ gives $g_\sigma(1,2)$;
it is written as
\begin{eqnarray}
&&V^\sigma_{\rm LDA}(\bbr) \!=\! V^\sigma_{\rm ext}(\bbr) + V^\sigma_{\rm Cou}(\bbr) + V^{\sigma}_{\rm LDA\_xc}(\bbr), \\
&&V_{\rm eff}^\sigma(\!\bribr2, \!\omega_p)\!=\! V_{\rm LDA}^\sigma(\bri) \delta(\bri\!-\!\br2)
            \!+\! \frac{\delta \OOSF}{\delta g^\sigma(\bribr2,\!\omega_p)}, \\
&&\left[ i \omega_p \!-\! \left(\frac{-\nabla_{\bri}^2}{2 m} \!+\! V_{\rm eff} \right)\right]
g^\sigma(\bribr2,\omega_p) \!=\!- \delta(\bribr2),
\label{eg2}
\end{eqnarray}
where $\sigma=\pm 1$ for $\up$ and $\dn$.
${\delta \OOSF}/{\delta g}$ is the contribution to the self-energy from SF.
$\omega_p$ denotes the fermion Matsubara frequencies.

It is difficult to include the full non-locality and
$\omega$-dependence of ${\delta \OOSF}/{\delta g}$.
So we restrict the variational space
as follows and try to find the optimum solution within it.
$V_{\rm LDA}$ gives exact solutions of $\Omega[g]$ if we omit $\OOSF$.
Therefore we take a variational space of $g_\sigma(1,2)$ so that it is generated
by $V_{\rm LDA}$ plus the onsite magnetic fields $B^R$
just acting on only $d$ electrons in each atomic-sphere (AS) $R$.
Thus $g_\sigma(1,2)$ within the variational space
is generated by solving a set of SC Eqs. for given $B^R$.
The set consists of Eq.(\ref{eg2}) and
\begin{eqnarray}
&&V_{\rm eff}^\sigma(\bribr2)\!=\! V_{\rm LDA}^\sigma(\bri) \delta(\bri - \br2)
  + \sigma \mu_{\rm B} \sum_R B_R \hat{P}_R.
\label{veff} 
\end{eqnarray}
Here $\hat{P}_R = \delta(r_1\!-\!r_2) \sum_m Y_{2m}(\theta_1,\phi_1) Y^*_{2m}(\theta_2,\phi_2)$
is the projection operator to the onsite $d$ channel.
Note that the set $\{B_R\}$ for all $R$
is only the variational parameters, which determine $g_\sigma(1,2)$.
Therefore, our problem is reduced to searching stationary point
of $\Omega[g]$ as a functional $\{B_R\}$.
(As in our static approximation $E_{\rm xc}[g] \approx E^{\rm LDA}_{\rm xc}[n]$,
the stationary point means minimum.)
This means that we have to solve
$ 0= {\delta \Omega/}{\delta g} \times {\delta g}/{\delta B_R}=
\left( -V_{\rm eff} + V_{\rm LDA} + {\delta \OOSF}/{\delta g} \right)
\times {\delta g}/{\delta B_R}$. This determines ${B_R}$
as the solution of an inversion equation
\begin{eqnarray}
\frac{\delta \OOSF}{\delta g} \frac{\delta g}{\delta B_R} =
\sum_{R'} \frac{\partial M_{R'}}{\partial B_R} B^{\rm SF}_{R'},
\label{invb}
\end{eqnarray}
where $M_R= \mu_{\rm B} \int_R d^3r (n^d_\uparrow - n^d_\downarrow)$.
$B_R = B^{\rm SF}_R + B^{\rm ext}_R$
if we add external magnetic field $B^{\rm ext}_R$ as a probe.
$M_R$ is the spin magnetic moment of $d$ electrons in $R$.
This kind of inversion equation like Eq.(\ref{invb}) is also used
in the optimized-effective-potential method \cite{kotani98}.
Equations (\ref{eg2}),(\ref{veff}), and (\ref{invb}) give a set of SC equations.
With the restriction of the variational space, all
the quantities in RHS in Eq.(\ref{omegag}) are treated as the functional
of $B_R$ in Eq.(\ref{veff}), or corresponding magnetic field $M_{R}$.

As for $\OSF[g]$, we give it in a simple model
in the onsite approximation;
it consists of contributions from each sites as $\OSF[g]=\sum_R \OSF_R[g_R]$.
Here $g_R$ denotes the $d$-channel onsite part of $g_\sigma(1,2)$ as
\begin{eqnarray}
&&g^\sigma_R({\bf r},{\bf r}',\omega_p)
= \sum_m g^\sigma_{mR}(\omega_p) \radphi(r,i\omega_p)  \nonumber\\&&\ \ \times
Y_{2m}(\theta,\phi) Y^*_{2m}(\theta',\phi') \radphi(r',i\omega_p).
\end{eqnarray}
Here we now take just the diagonal part with respect
to the magnetic quantum number $m$.
$\radphi(r,i\omega_p)$ denotes solutions of the radial Schr\"odinger equation.
$\OSF_R[g_R]$ is given as
\begin{eqnarray}
&&\OSF_R[g_R]
= \frac{1}{\beta} \sum_{\nu_q}
  Q_0^{\rm RPA}(I_R^{\rm L} \chi^{\rm 0L}_R(\nu_q)) \nonumber \\  \ \ &&
+ \frac{2}{\beta} \sum_{\nu_q}
  Q_0^{\rm RPA}(I_R^{\rm T} \chi^{\rm 0T}_R(\nu_q)), \label{omgsf} \\
&&  Q_0^{\rm RPA} (s) \equiv \log (1-s) + s \label{qrpa}
\end{eqnarray}
$\nu_q$ denotes the bosonic Matsubara frequencies.
The bare longitudinal and transversal onsite spin-polarization functions,
$\chi_R^{\rm 0L}(\nu_q) =
\langle \hat{S}_{zR} \hat{S}_{zR} \rangle_0$ and
$\chi_R^{\rm 0T}(\nu_q)= \frac{1}{2}
\langle \hat{S}_{xR} \hat{S}_{xR} +\hat{S}_{yR} \hat{S}_{yR} \rangle_0$,
are given as
\begin{eqnarray}
&&\chi_R^{\rm 0L}(\nu_q) = \frac{-1}{2 \beta}
\sum_m \sum_{\omega_p} \big\{
g^\up_{mR}(\omega_{p}) g^\up_{mR}(\nu_q+\omega_{p}) \nonumber \\ &&\hspace{1cm}
+g^\dn_{mR}(\omega_{p}) g^\dn_{mR}(\nu_q+\omega_{p})\big\} W_{mR}(i\omega_p,i\nu_q),
\label{chi0l}\\
&&\chi_R^{\rm 0T}(\nu_q) =
\frac{-1}{2 \beta}
\sum_m \sum_{\omega_p} \big\{
g^\up_{mR}(\omega_{p}) g^\dn_{mR}(\nu_n+\omega_{p}) \nonumber \\ &&\hspace{1cm}
+g^\dn_{mR}(\omega_{p}) g^\up_{mR}(\nu_n+\omega_{p})\big\} W_{mR}(i\omega_p,i\nu_q),
\label{chi0t}\\
&&W_{mR}(i\omega_p,\! i\nu_q) \!=\! \left\{ \int^{R_{\rm max}}_0 
\hspace{-.7cm} r^2 dr \Phi_{mR}(r,\!i\omega_p) \Phi_{mR}(r,\!i\nu_q\!+\!i\omega_p) 
\right\}^2\!\!,
\end{eqnarray}
where $W_{mR}$ is the square of the radial part of integrals.
${R_{\rm max}}$ denotes the radius of AS $R$.
$\OSF_R[g_R]$ is the SF energy given by the onsite-only RPA,
though we treat the effective interactions
$I_R^{\rm L}$ and $I_R^{\rm T}$ as the functionals of $B_R$.
They are determined in a SC manner explained later.
Together with $\OSFRzero$ calculated in the same procedure
with taking the $T \to 0$ limit in Eq.(\ref{omgsf}),
we can obtain $\OOSF[g]=\sum_R \OOSFR$, where $\OOSFR \equiv \OSF_R - \OSFRzero$.
Corresponding to our onsite approximation on $\OOSF[g]$,
we only take diagonal terms on $R$ in the inversion Eq.(\ref{invb}).
That is, we just take ${\delta g_R}/{\delta B_R}$, and ${\partial M_R}/{\partial B_R}$,
neglecting the off site contributions.
Then Eq.(\ref{invb}) is simplified as
\begin{eqnarray}
B^{\rm SF}_{R} = \left(\frac{\delta \OOSFR}{\delta g_R} \frac{\delta g_R}{\delta B_R} \right)
\left(\frac{\partial M_{R}}{\partial B_R}\right)^{-1}.
\end{eqnarray}

In practice, $\chi^{\rm 0}_R(\nu_q)$,
which denotes each term contained in
$\chi_R^{\rm 0L}$ and $\chi_R^{\rm 0T}$ of Eqs.(\ref{chi0l},\ref{chi0t}),
is calculated through these formulas symbolically written as
\begin{eqnarray}
&&\chi^{\rm 0}_R(\nu_q) = \int_{-\infty}^{\infty} \frac{\Gamma^{0}(\nu) d \nu}{i \nu_q - \nu},
\label{convol0} \\
&&\Gamma^{0}(\nu) = \int_{-\infty}^{\infty} \hspace{-.5cm}d \omega
\left\{ f(\omega)\!-\!f(\nu\!+\!\omega) \right\} X(\nu\!+\!\omega) Y(\omega) W_{mR}(\omega,\nu).
\label{convol}
\end{eqnarray}
Here $f(\omega)$ in this convolution Eq.(\ref{convol})
denotes the Fermi distribution function;
$X(\omega),Y(\omega)$ denotes the partial density of states (DOS) $D^\sigma_{mR}(\omega)
\!=\! -{\rm Im} g^\sigma_{mR}(\omega)/\pi$. The factor $W_{mR}(\omega,\nu)$
introduces a natural cutoff for the convolution of DOS in Eq.(\ref{convol}).

It is necessary to include contributions due to
$\langle \hat{S}^{m}_{zR} \hat{S}^{m}_{zR} \rangle_0$
which carries SF with other magnetic angular momentum $m\ne 0$
in addition to the contribution of $m=0$
($\hat{S}^{m=0}_{zR}$ is $\hat{S}_{zR}$ given above).
If we assume spherical symmetry,
we can evaluate it as $\langle \hat{S}^{m}_{zR} \hat{S}^{m}_{zR} \rangle_0
\approx \frac{5-|m|}{5} \langle \hat{S}_{zR} \hat{S}_{zR} \rangle_0$
by counting the number of allowed combinations of $m_1$ and $m_2$ for $m=m_1+m_2$;
Then we can include the contributions by using
\begin{eqnarray}
Q^{\rm RPA}(s) \equiv \sum_{m=-4,...4} Q^{\rm RPA}_0\left(\frac{5-|m|}{5} s\right)
\end{eqnarray}
instead of $Q^{\rm RPA}_0(s)$ in Eq.(\ref{qrpa}).
Our system is not spherically symmetric but we utilize
this just as a convenient method in order to evaluate
$\OSF_R[g_R]$ only through quantities
$\chi_R^{\rm 0L}$ and $\chi_R^{\rm 0T}$
in addition to $I_R^{\rm L}$ and $I_R^{\rm T}$.

A key point in our method is in the SC
determination of $I_R^{\rm L}$ and $I_R^{\rm T}$.
Here we only treat the paramagnetic ground state at finite-temperature.
At first, let us consider the case without $\overline{\Omega^{\rm SF}}[g]$;
this means that $\Omega[g]$ in Eq.(\ref{omegag}) is reduced to be
the crude finite-temperature LDA.
Then $D^\sigma_{mR}(\omega)$ in each site $R$
is given as a functional of
$B^{\rm ext}_{R}$ as the solution of an impurity problem;
adding the perturbation $B^{\rm ext}_{R}$ only at a site $R$.
This problem is easily treated in AkaiKKR.
Then $\chi^{\rm 0L}_R(0)$ is calculated
through Eqs.(\ref{convol0},\ref{convol})
from $D^\sigma_{mR}$ as a function of $B^{\rm ext}_R$.
On the other hand, we can also calculate
the longitudinal local static spin susceptibility at the site $R$
from the numerical 2nd derivative of $\Omega[g]$
with respect to $B^{\rm ext}_R$ as
$\displaystyle \chi_R^{\rm L}(0)
=\langle \hat{S}_{zR} \hat{S}_{zR} \rangle
= \frac{1}{4\mu_{\rm B}^2} \frac{\partial M_R}{\partial B^{\rm ext}_R}$.
Note that these $\chi^{\rm 0L}_R(0)$ and $\chi^{\rm L}_R(0)$
are calculated as functions of $B^{\rm ext}_R$.
At the same time, we also calculate
$\displaystyle \chi^{\rm T}_R(0) = \frac{1}{4\mu_{\rm B}^2}
\frac{M_R}{ B^{\rm ext}_R}$ and $\chi^{\rm 0T}_R(0)$;
we evaluate $\chi^{\rm T}_R(0)$
as if our system is spherical symmetric for simplicity.
Then we can calculate $I_R^{\rm 0L}$ so that
$\left\{ \chi_R^{\rm L}(0) \right\}^{-1}
= \left\{\chi_R^{\rm 0L}(0)\right\}^{-1} - I_R^{\rm 0L}$.
With this $I_R^{\rm 0L}$, we can calculate the longitudinal part
of $\OSF_R$. The same method is used also for $I_R^{\rm 0T}$ so as to be
$\left\{ \chi_R^{\rm T}(0) \right\}^{-1}
= \left\{ \chi_R^{\rm 0T}(0)\right\}^{-1} - I_R^{\rm 0T}$.
As for $\OSFRzero$,
we can calculate it in the same manner above from
the same $D^\sigma_{mR}(\omega)$, $\chi^{\rm L}_R(0)$,
and $\chi_R^{\rm T}(0)$,
but taking the $T \to 0$ limit in Eq.(\ref{omgsf})
through the Matsubara frequencies.
As a result, we can obtain $\OOSFR(M_R)$ as a function of $M_R$.
This $\OOSFR(M_R)$ should give an effect to make the system
paramagnetic above $T_{\rm c}$.
We should have to include the effect
from the beginning, that is, we rather have to start from
$\Omega[g]$ including $\OOSF[g]= \sum_R \OOSFR(M_R)$.
This means that we can obtain a new function $\OOSFR(M_R)$
if we start from $\Omega[g]$ with a trial function $\OOSFR(M_R)$.
We impose a self-consistency condition, namely the agreement
between these functions. In practice, we assume a simple form
$\OSF_R = \OSF_0 + \frac{1}{2} \alpha^{\rm SF}(M_R/\mu_{\rm B})^2$.
Then our self-consistency is just for the parameter $\alpha^{\rm SF}$,
as the trial $\alpha_{\rm in}^{\rm SF}$ and
the calculated $\alpha_{\rm out}^{\rm SF}$ are the same.
This assumption is necessary
so as to make the SC equation determining $\OSF_R(M_R)$ be closed.
We can take the self-consistency used here
as a simplified version of the general self-consistency scheme
in the construction of the LW functional;
the 2nd derivative of $\Omega[g]$ with respect to $g_\sigma(1,2)$
can give the fluctuation (or 2-body propagators);
on the other hand, the fluctuation can be used to construct
$\Omega_{\rm xc}[g]$ through the coupling-constant-integral formula
\cite{kotani98}.
In the STLS theory \cite{stls} for
homogeneous electron gas and the SCR theory \cite{moriya},
other simplified versions of the general scheme were used.
\begin{figure}
\begin{center}
\caption{Determination of $\alpha^{\rm SF}$:
This is an example for Fe.
We determine $\alpha^{\rm SF}$ so that
$\alpha_{\rm in}^{\rm SF}=\alpha_{\rm out}^{\rm SF}$.See text.}
\mbox{\epsfig{file=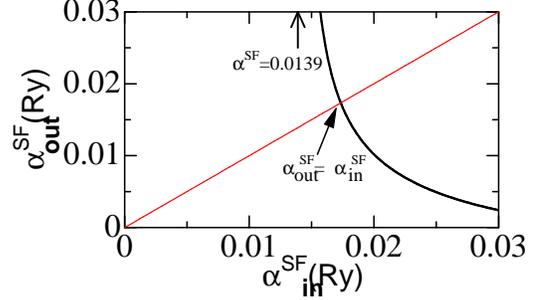,height=4cm}}
\label{alpha}
\end{center}
\end{figure}
In Fig.\ref{alpha}, we give an example of plots of
$\alpha_{\rm in}^{\rm SF}$ vs. $\alpha_{\rm out}^{\rm SF}$,
determining $\alpha^{\rm SF}$.
In our formalism, the interaction
$I_R^{\rm 0L}$ has a part that is linearly-dependent on $\alpha_{\rm in}^{\rm SF}$.
The most divergent part in Eq.(\ref{omgsf}) occurs at $\nu_0=0$ in
$\log(1-I_R^{\rm 0L} \chi_R^{\rm 0L}(0))=
\log\left(\left\{\chi_R^{\rm L}(0)\right\}^{-1} \chi_R^{\rm 0L}(0) \right)$;
$\log\left(\chi_R^{\rm L}(0) \right)
=-\log\left(\overline{\chi_R^{\rm L}(0)} +4\alpha_{\rm in}^{\rm SF}\right)$,
where $\overline{\chi_R^{\rm L}(0)}$ denotes $\chi_R^{\rm L}(0)$
at $\alpha_{\rm in}^{\rm SF}=0$.
It diverges at $\alpha_{\rm in}^{\rm SF} \to
-\frac{1}{4} \left\{\overline{\chi_R^{\rm L}(0)}\right\}^{-1}$.
This occurs at
$\sim \alpha_{\rm in}^{\rm SF}=0.0139$ Ry denoted
with the arrow in Fig.\ref{alpha}.
In the case of Fe shown in Fig.\ref{alpha} and fcc-Co,
$-\left\{\overline{\chi_R^{\rm L}(0)}\right\}^{-1}$
is positive even at $T=0$ because
LDA gives a local static magnetic moment \cite{localm}.
Therefore our treatment gives
$\alpha^{\rm SF} \to -\frac{1}{4} \left\{\overline{\chi_R^{\rm L}(0)}\right\}^{-1}>0$
at $T \to 0$, thus $\chi_R^{\rm L}(0) \to \infty$.
Thus $\OSF$ keeps $\chi_R^{\rm L}(0)>0$ and
forbids the existence of the static local magnetic moment at $T = 0$.
This is physically reasonable as a ground state character, though
we still have an unphysical divergence at $T\to 0$.
In these cases, the results does not coincide with the LDA limit.
On the other hand, as for Ni and Pd, which show negative
$-\left\{\overline{\chi_R^{\rm L}(0)}\right\}^{-1}$, it coincides
with the LDA limit with $\alpha^{\rm SF} \to 0$ at $T\to 0$.
\newpage
\begin{figure}
\begin{center}
\caption{ Top Panel: Bulk Magnetic susceptibility $\chi^{-1}$.
The middle panel: The local onsite susceptibilities
$\{ \chi_R^{\rm L}(0) \}^{-1}, \{\chi_R^{\rm 0L}(0)\}^{-1}$, and
the effective onsite interaction
$I_R^{\rm L}=\{ \chi_R^{\rm L}(0) \}^{-1}-\{\chi_R^{\rm 0L}(0)\}^{-1}$.
The loser panel: $\alpha^{\rm SF}$ in
$\OSF_R = \OSF_0 + \frac{1}{2} \alpha^{\rm SF}(M_R/\mu_{\rm B})^2$.
Solids lines for experimental lattice constants,
broken lines for 3\% smaller lattice constants. See text.}
\mbox{\epsfig{file=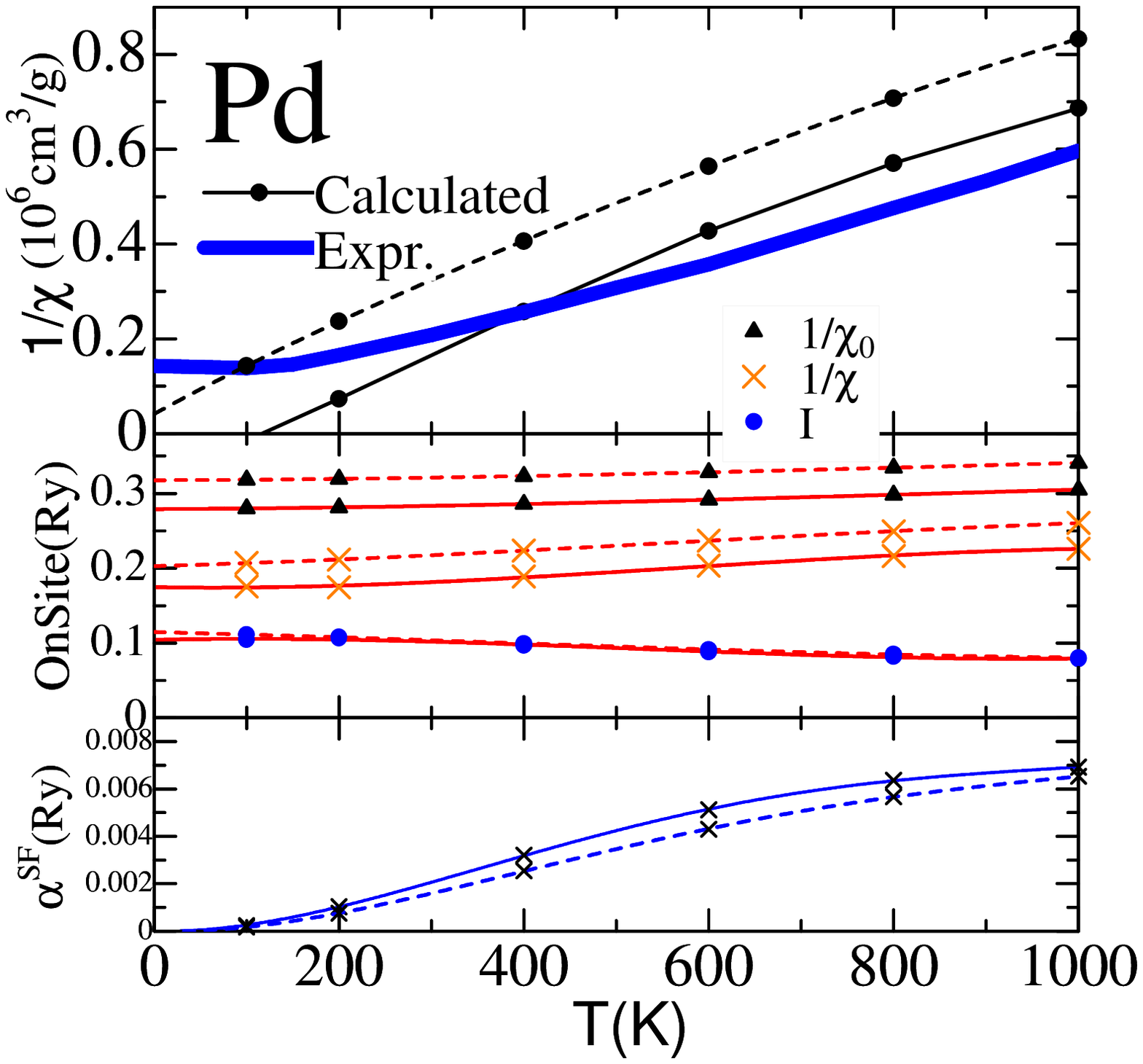,height=8cm}}
\mbox{\epsfig{file=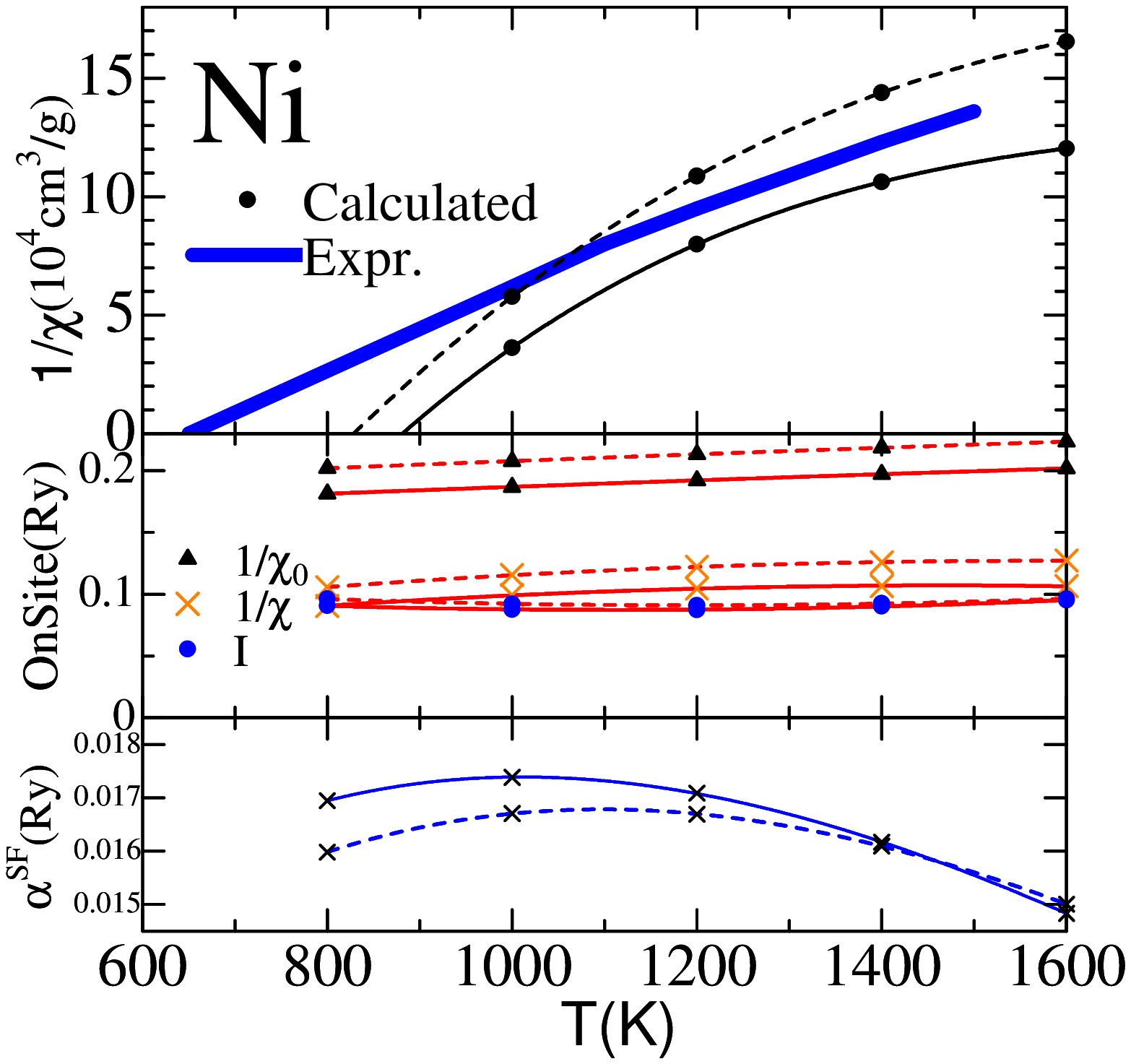,height=8cm}}
\newpage
\mbox{\epsfig{file=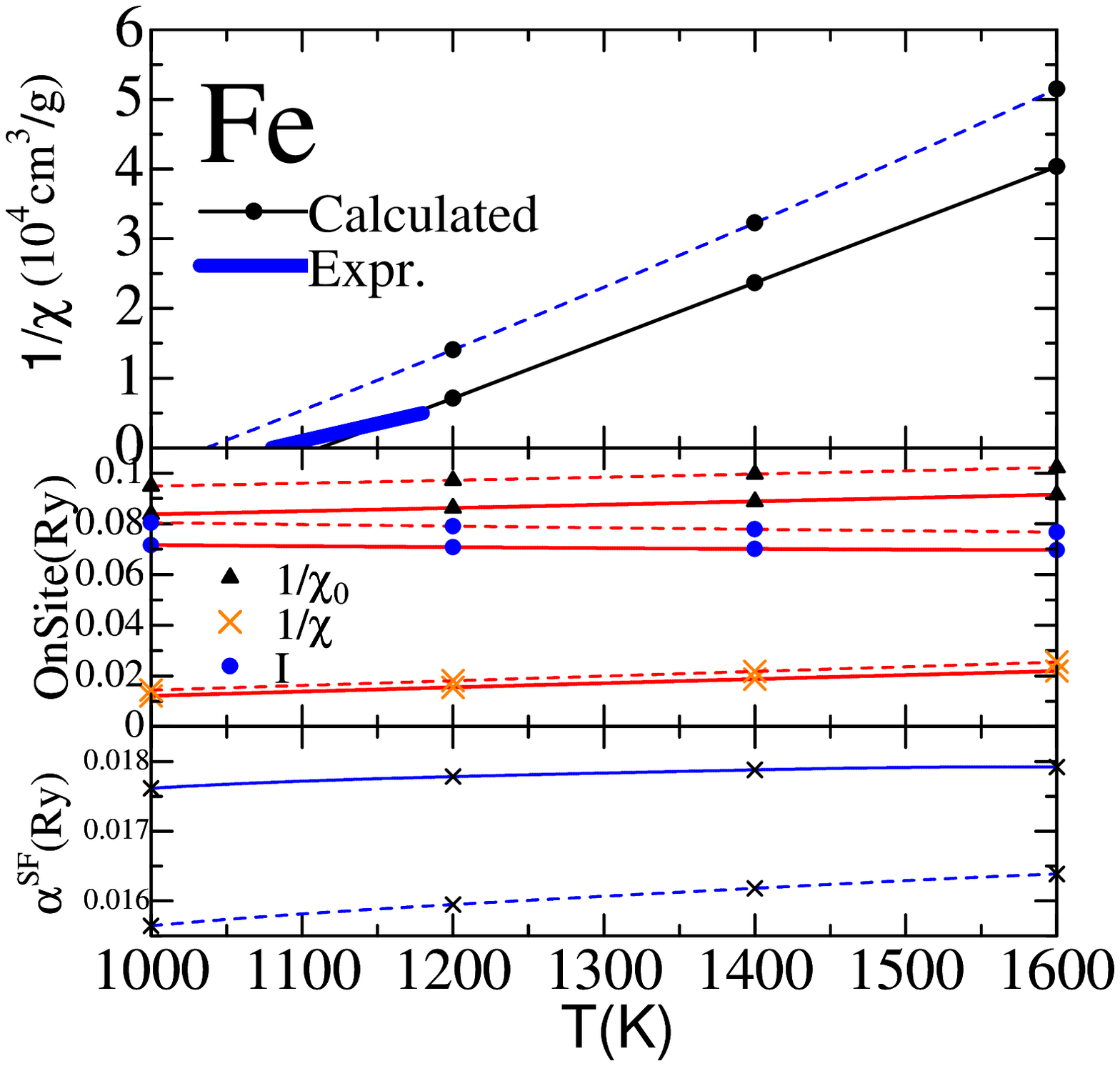,height=8cm}}
\mbox{\epsfig{file=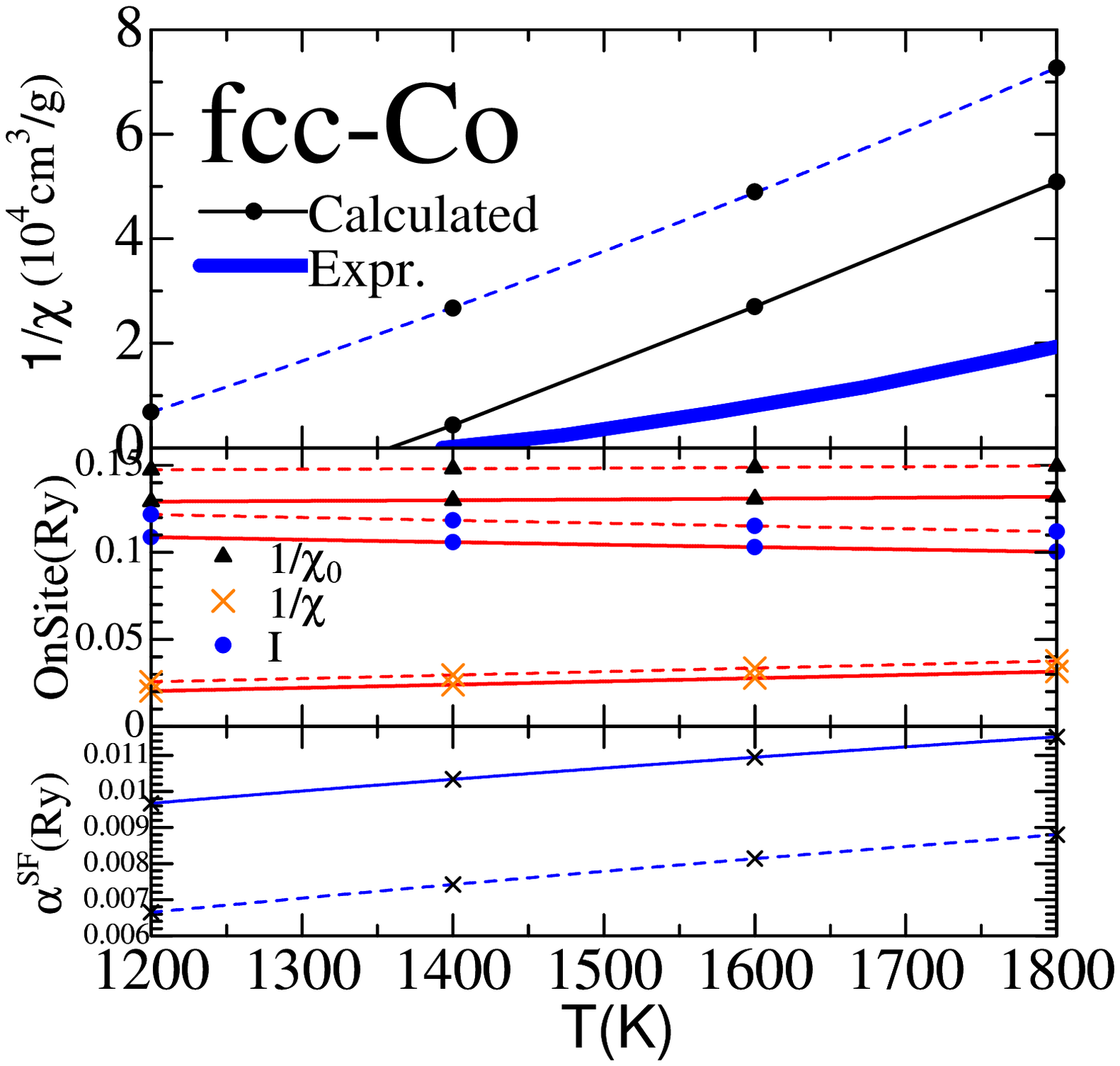,height=8cm}}
\label{chi}
\end{center}
\end{figure}
The partial DOS $D^\sigma_{mR}(\omega)$ are calculated
with $\sim$ 8000 $k$ point in the irreducible Brillouin zone (BZ),
where each poles of eigenvalues are effectively
smeared with Lorentzian functions
$\propto 1/(\omega^2 + \Delta_{\rm L}^2)$.
We needed to perform calculations
with $\Delta_{\rm L}=0.009,0.012,0.016,0.022$Ry cases
and to take an extrapolation $\Delta_{\rm L} \! \to \! 0$.
After we obtain the self-consistently determined $\alpha^{\rm SF}$,
we can calculate the bulk susceptibility $\chi$ for the whole systems.
In Fig.\ref{chi}, we summarise our results.
Used lattice constants $a=$
5.47, 6.70, and 6.80 a.u. for bcc Fe, fcc Ni, and fcc-Co
(experimental values at $T_c$);
$a=$7.35 a.u. for fcc Pd (experimental value at $T=0$).
They are shown with solid lines.
In addition, we show the cases for 3\% smaller lattice constants
by the broken lines; the differences from solid lines
indicate errors due to LDA because the ordinary LDA calculations
give rise to errors of this level in the lattice constants.
The bottom panels show $\alpha^{\rm SF}$.
The top panels are for the inverse of
the bulk magnetic susceptibilities $\chi^{-1}$
calculated from these $\alpha^{\rm SF}$,
where we also show experimental results taken from Ref.\cite{lb}.
The middle panels are for
$\left\{\chi_R^{\rm 0L}(0)\right\}^{-1}$,$\left\{\chi_R^{\rm L}(0)\right\}^{-1}$
and $I=\left\{\chi_R^{\rm 0L}(0)\right\}^{-1}-\left\{\chi_R^{\rm L}(0)\right\}^{-1}$.

As for Pd, our LDA calculations without $\OOSF$ in ASA at $T=0$
give ferromagnetism for $a$=7.35 a.u. and paramagnetism for $a=7.13$ a.u.
In Ref.\cite{staunton2000}, the LDA calculation with $a=7.35$
gives paramagnetic phase. This discrepancy might arise
from the difference in the formalism and details of
implementation \cite{diffakaikkr}.
However, the difference seems not important
for the overall discussions here.
The temperature dependence of $\chi^{-1}$, especially slope
at a high temperature region,
is rather well described in spite of no tunable parameters.
It fails to describe $\chi^{-1}$ around $T=0$
where we expect only SF around ${\bf k}\sim 0$
dominates the behaviour.
It is reasonable because our formalism could work well
only in the case that SF in each site behaves independently; in other words,
all the modes of SF are excited rather homogeneously in BZ.
As $T$ becomes higher, $\alpha^{\rm SF}$ increases smoothly,
which suppresses SF at ${\bf k}=0$ ($\chi^{-1}$ gets larger).
On the other hand, the local SF $\chi_R^{\rm 0L}(0)$ does not
change much. The enhancement factor on local SF
$\chi_R^{\rm L}(0)/\chi_R^{\rm 0L}(0)$
due to the interaction $I_R^{\rm L}$, is not so large
$\sim 1.5$ at $T=0$ and gradually decreases for larger $T$.
In Ni, our method can also describe experimental
$T_c$ and the slope of $\chi^{-1}$ reasonably.
$\alpha^{\rm SF}$ decreases above $T \sim 1000$K.
$\chi^{-1}$ is not so linear as experiments and
contains negative-quadratic dependence
as $\chi^{-1} \propto -T^2$
reflecting the behaviour of $\alpha^{\rm SF}$.
In both cases of Pd and Ni, it seems that
we have to add positive-quadratic dependence $\chi^{-1} \propto T^2$,
so as to have better agreement with experiments.
We might expect such an effect if we include
${\bf k}$-dependent SF,
because then we will have larger contributions even for lower temperatures.

As for Fe and fcc-Co, calculated $\chi^{-1}$ is rather linear.
Slopes of $\chi^{-1}$ are too large compared with experiments,
though calculated $T_{\rm c}$'s are still reasonable.
The factors $\chi_R^{\rm L}(0)/\chi_R^{\rm 0L}(0)$ are rather
large $\sim 5$ in these cases.
The relation $\left\{\chi_R^{\rm L}(0)\right\}^{-1} \to 0$ at $T\to 0$
should be hold in agreement with the previous discussion.
This condition keeps the temperature dependence of
$\left\{\chi_R^{\rm L}(0)\right\}^{-1}$ too large,
corresponding to too large a slope of $\chi^{-1}$.
As is same the as cases in Ni and Pd, we expect better agreements
by taking account of ${\bf k}$-dependent SF
which will remove the above condition from our formalism.

We have shown results of paramagnetic susceptibilities for some metals.
They give reasonable agreements with experiments, 
indicating the possibilities of treatments 
along the present line including SF self-consistently.
In particular, the results seems to indicate necessity
of including the ${\bf k}$-dependence of SF for the further improvement.

We thank Dr.Aryasetiawan for valuable discussions and
comments on the manuscript.




\begin{references}
\bibitem{diffakaikkr}
Our KKR-ASA contains full energy-dependence of radial functions
and including the energy dependence of eigenfunctions in the interstitial region.
\bibitem{localm}
Ni atom, which is treated as an impurity,
in paramagnetic Ni matrix does not show magnetic moment
in the self-consistent impurity calculation in LDA.
The same is also fro Pd. On the other hand, Fe and fcc-Co shows
"local moment" in such a calculation.
\end{references}
\end{document}